\documentclass[12pt]{article}
\usepackage{dina4p}
\usepackage{amsfonts}
\usepackage{graphicx}

\begin{document}

\title{Spin motion at and near orbital resonance 
       in storage rings with Siberian Snakes \\
       I: at orbital resonance\thanks{ DESY preprint DESY 06--220.
       Published in: {\sl New Journal of Physics} {\bf 8} (2006) 296. 
       Worldwide copyright by:  Institute of Physics and Deutsche Physikalische Gesellschaft (2006)}
       }

\author{ D~P~Barber\thanks{Also Visiting Staff Member at the Cockcroft Institute, Daresbury
                             Science and Innovation Campus, and at the
                             University of Liverpool, UK.}  \\ {\tt mpybar@mail.desy.de}
    \and M Vogt \\ {\tt vogtm@mail.desy.de}
        }
\date{Deutsches~Elektronen--Synchrotron, DESY, ~22607 ~Hamburg, ~Germany}

\maketitle


\begin{abstract} 
  
  Here, and in a sequel, we invoke the invariant spin field to provide an in--depth study of
  spin motion at and near low order orbital resonances in a simple
  model for the effects of vertical betatron motion in a storage ring
  with Siberian Snakes.  This leads to a clear understanding, within
  the model, of the behaviour of the beam polarisation at and near 
  so--called snake resonances in proton storage rings.

\end{abstract}

\section{Introduction}

In earlier papers we and collaborators have emphasised the utility of
the {\em invariant spin field} (ISF) and the {\em amplitude dependent
  spin tune} (ADST) for analysing spin motion in circular particle
accelerators and storage rings
\cite{hh96,hvb99,gh2006,mv2000,spin2002,hv2004,beh2004,behresponse,hde2006,ky88}. 
In particular,
under certain conditions, the ISF is unique up to a global  sign and in that case it allows
estimates to be made of the maximum equilibrium beam polarisation and the
maximum time averaged beam polarisation in proton storage rings.  Then, for example, 
for a given equilibrium distribution of particles in phase space,
the maximum attainable polarisation at the chosen high energy can
be estimated before embarking on  extensive  computer simulations of the effect on the
polarisation of acceleration from low energy.  Once a 
machine configuration has been found which appears to be acceptable 
at the chosen high energy, one
then studies the effect of acceleration to assess whether the configuration
is still acceptable. Acceleration can involve crossing
many {\em spin--orbit resonances} and that can lead to a loss of
polarisation.  The latter problem can be partially solved by the
inclusion in the ring of so--called {\em Siberian Snakes} \cite{dk76,dk78}, magnetic
field configurations that cause the average spin precession rate on the design
orbit to be independent of the nominal beam energy. Nevertheless,
spin--orbit resonances can still occur but their identification then
often requires a more careful definition of the spin precession rate
than has been common among practitioners, involving  the
amplitude dependent spin tune. A full
understanding also requires a careful definition of an adiabatic
invariant for spin motion.  In most of the numerical investigations described in
\cite{hh96,hvb99,gh2006,mv2000,spin2002,hv2004,beh2004,hde2006,ky88}, 
orbital resonance is avoided. Moreover, it is
shown that the spin--orbit systems tend to avoid exact spin--orbit
resonance.
These and other matters are explained and
illustrated in great detail in the sources cited above. In order to
keep this paper to a reasonable length we will assume that the reader
is familiar with that material.

Of course, the ISF and the concepts derived from it, may be of little
help if the ISF is not unique. That can be the case if the orbital
motion is resonant or if the system is on spin--orbit resonance
\cite{spin2002,beh2004}.  Nevertheless, as we show below, special choices
from sets of non--unique ISF's can be useful for investigating spin
motion near some kinds of orbital resonance. Moreover, the ISF is
still useful at rational vertical orbital tunes corresponding to the
so--called odd order snake ``resonances''. At these tunes the Siberian
Snakes apparently  do not succeed in preventing loss of polarisation during
acceleration \cite{lt86,syl88,sylbook,luccio1995,ptitsyn2004,mbai2004}.  However, 
with the exceptions of \cite{ptitsin97,ptitsin95,spin2002,mane2004}, discussions
about spin motion at or near to these tunes have made no reference to the ISF. 
The treatment in \cite{ptitsin97,ptitsin95} involved a mathematical approximation to
the model used in this paper.
Then in \cite{spin2002} it was pointed out for the first
time that at these tunes the ISF is an irreducibly discontinuous
function of the vertical orbital phase and that the discontinuities
can be moved, thereby demonstrating non--uniqueness.  
In \cite{mane2004} the necessity of the discontinuities was disputed (see Section 3.4).
In Section 2 we explain that exactly at these special tunes, the term ``snake resonance'' 
does not fit with our preferred definition of spin--orbit resonance. Nevertheless,
for simplicity, we adopt the now traditional nomenclature.
In \cite{spin2002} it was also made clear how non--uniqueness can occur at other rational tunes.

In this paper and in a sequel (called Part II) we extend the
investigations in \cite{spin2002}.  In the initial and pioneering work
on snake resonances in \cite{lt86,syl88,sylbook}, emphasis was
placed on the significance of the so--called ``perturbed spin tune'',
a measure of the angles of spin rotation around the real, unit length,
eigenvectors of 1--turn SO(3) spin maps. See also \cite{leemane}.  However,
these eigenvectors are usually not solutions of the Thomas--Bargmann--Michel--Telegdi (T--BMT) equation
along the trajectories. Thus, while it is clear from calculations that the
``perturbed spin tune'' can show strong variations, we do not consider
its behaviour to be relevant to the discussion \cite{behresponse}.
In \cite{lt86,syl88,sylbook} spin motion was also analysed in terms of
an essentially perturbative expansion of the $p$--turn SU(2) spin
transfer matrix, $T(p)$, and it was found that at snake--resonance
tunes, $|T_{2 1}(p)|$ could increase without limit as the number of
turns $p$ increased. In so far as it relates to positions in tune space,  
this behaviour, which is an artifact of the perturbative
approach, appears to be consistent with
the snake resonance phenomenon. 
However, although an unlimited increase of a matrix element in a
perturbative expression for a rotation matrix does suggest exceptional
behaviour, it destroys the unitarity of the matrix, thereby
demonstrating an invalid approximation and implying a consequent
limitation of the predictive power of the calculation.
For example, in the absence of other input, one might suppose  that 
an unlimited growth of $|T_{2 1}|$ could infer that initially vertical spins 
are simply flipped. Alternatively, the growth might be a hint  
that the vertical component of the beam polarisation oscillates
as  spins rotate around a horizontal axis.
Lastly, simulations reported in \cite{ky88,buon85,ptitsin952} demonstrate 
the effects of  varying the rate of acceleration near snake--resonance tunes.
The number of turns needed to traverse a given energy
range depends on the energy gain per turn. Then, if weight is given to the
perturbative treatment,  the number of turns
determines how large  $|T_{2 1}|$ can become.
The rate of acceleration is certainly important in the
Froissart--Stora calculation \cite{fs60} of the loss of polarisation  
when crossing spin--orbit resonances in rings without snakes, and the 
phenomenology is well understood.
However, the simulations in \cite{ky88,buon85,ptitsin952} show that the dependence 
of the final polarisation on the
acceleration rate can be complex and unexpected and that no clear picture emerges.

To summarise, in our opinion, although snake resonances have presented
problems \cite{ptitsyn2004,mbai2004}, the numerical and
theoretical investigations made so far have provided no completely 
coherent picture of spin motion at and near 
snake--resonance tunes, either with or without acceleration.

These papers provide a new contribution towards such a picture,
at least within our adopted simple model.
We carry out our study against the background of our
standard philosophy, namely that to detect exceptional behaviour, one
should start spin--orbit tracking simulations with an equilibrium
distribution of particles in phase space and with each spin parallel
to the ISF vector corresponding to the position of the particle in
phase space \cite{beh2004}.  Then any unexpected behaviour is
signalled by long term or turn--to--turn variations of the
polarisation of the beam.  This gives a much cleaner view of the
situation than if one just begins in the common way with spins
parallel to the direction of the ISF on the closed orbit. 
Accordingly, with the ISF at the centre of our
discussion we show how, in the cases considered, the long term
behaviour of spins can be inferred, at least qualitatively, from some
features of the ISF. On low order
orbital resonance, an ISF can be calculated almost trivially from the
spin maps of a few turns.

 For our purposes, and in order to allow direct comparison, it suffices
just to consider a model used in earlier literature \cite{lt86,syl88,sylbook},
namely a model with two Siberian Snakes.  Since the ranges of the
relevant parameters and the number of possible configurations is huge,
this study, which is mainly numerical, is not exhaustive. We fully
appreciate that storage rings do not run on low order orbital
resonance, that spin--orbit resonances need not be well separated, 
that particles have three modes of oscillation and that
particle motion in real rings can be nonintegrable. Nevertheless our
study provides useful insights.

The paper is structured as follows.  We continue in Section 2 by
recalling the simple idealised and traditional model of spin motion
for protons considered in \cite{spin2002,lt86,syl88,sylbook} and specify the
notation commonly used to describe it.  Then in Section 3 we use the
model to study spin motion exactly at orbital resonances including an
odd order snake resonance and show how the chief features of spin
motion can be guessed from the characteristics of the ISF.
We summarise our studies in Section 4. Part II of this study 
completes the picture by addressing  spin motion close to, but not at,  an 
odd order snake--resonance tune. 
The numerical calculations were carried out with purpose--built spin--orbit 
tracking codes, with the spin--orbit 
tracking facilities in the code SPRINT \cite{gh2006,mv2000} and with
the SODOM--II algorithm \cite{ky99} embedded in SPRINT.

\section{Recapitulation -- the single resonance model with two snakes}

Spin motion in the electric and magnetic fields at the point $\vec z$ in the 6--dimensional
phase space at beam energy $E_0$ and at the position $s$ around the ring, is
described by the T--BMT precession equation $d \vec{S}/{d s}= {\vec \Omega}
(\vec z; s, E_0) \times \vec {S}$ ~\cite{jackson,bhr1,hh96} where $\vec S$ is
the spin expectation value (``the spin'') in the rest frame of the
particle and ${\vec \Omega}(\vec z; s, E_0)$ contains the electric and magnetic
fields in the laboratory and depends on the beam energy $E_0$.  The ISF, whose value at $(\vec z; s)$
is denoted by $\hat n(\vec z; s)$, is
a 3--vector field of unit length obeying the T--BMT equation along
particle trajectories $(\vec z(s); s)$ and fulfilling the periodicity
condition ${\hat n}(\vec z; s + C) = {\hat n}(\vec z; s)$ where $C$ is
the circumference\footnote{We emphasise that the non--trivial ISF  vector $\hat n(\vec z; s)$ 
    should not be confused with the trivial vector $\vec n$ used to denote $\vec \Omega$ in \cite[equation 2.46]{sylbook}
    and in \cite[equation 1]{syl2006} and having the same periodicity.}.
Thus ${\hat n}({\vec M}(\vec z; s); s + C) = {\hat
  n}({\vec M}(\vec z; s); s) = R_{_{3 \times 3}}(\vec z; s) {\hat
  n}(\vec z; s)$ where ${\vec M}(\vec z; s)$ is the new position in phase space
after one turn starting at $\vec z$ and $s$, and $R_{_{3 \times
    3}}(\vec z; s)$ is the corresponding spin transfer matrix.  
For convenience we have suppressed the dependence of $\vec M, R$ and ${\hat n}$ on $E_0$.
In addition to the kinematical constraint $|\hat n| =1$, a complete
definition of the ISF requires the specification of a constraint on
its regularity with respect to $\vec z$. For example, one could
require that $\hat n (\vec z; s)$ is continuous in $\vec z$.  It is
clear that such regularity conditions are needed since, for example, a
piece--wise continuous ISF exists if a continuous one exists but not
vice versa. See Section 3.4 and \cite{spin2002}. However, since the
emphasis of the paper is on numerical results, we only occasionally
dwell on the matter of regularity. We use the term ``global
uniqueness'' if two ISF's can differ only by a sign. Thus in
the case of global uniqueness, either exactly two ISF's, $\pm \hat n$,
exist as in Section 3.1 or none, as in Section 3.4.  We use the term
``local uniqueness'' if any two ISF's, $\hat n$ and $\hat n'$ are
parallel, i.e. $\hat n \times \hat n' =0$, so that $\hat n$ and $\hat
n'$ can differ only by a sign function. Of course global uniqueness
implies local uniqueness but not vice versa. Since the issue of local
uniqueness is beyond the scope of this paper, it will be addressed
only briefly.  If an ISF exists and parameters such as $E_0$ are constant, the scalar product $J_{\rm s} = \vec
S \cdot \hat n/|\vec S|$ is invariant along a trajectory.

For a turn--to--turn invariant particle distribution in phase space, a
distribution of spins initially aligned along the ISF remains
invariant from turn--to--turn, i.e., in ``equilibrium''. Moreover, for
integrable orbital motion and away from both orbital resonances and 
spin--orbit resonances (see below), the average $|{\langle\hat
n(\vec z; s)\rangle}|$ of $\hat n$ over the phases on a torus is the
maximum attainable time averaged beam polarisation $P_{_{\rm lim}}$.  
Away from orbital resonances and spin--orbit resonances the actual time 
averaged polarisation can be written as  $P_{_{\rm lim}} P_{_{\rm dyn}}$ 
where the $P_{_{\rm dyn}} = |\langle J_s \rangle|$ depends on the history 
of the beam \cite{mv2000}. For a turn--to--turn invariant particle distribution in phase space
$P_{_{\rm lim}} = |{\langle\hat n(\vec z; s)\rangle}|$
is also the maximum attainable equilibrium beam polarisation. This is reached when $P_{_{\rm dyn}} =1$. 

Under appropriate conditions $J_{\rm s}$ is an adiabatic invariant
while system parameters such as the beam energy $E_0$ are slowly varied
\cite{gh2006,hde2006}. In fact $\hat n$  then serves as a ``template''
for spin motion. Several examples of this are given in Section 3.

The ADST ${\nu}_{{\rm s}}(\vec J)$ at the amplitudes (actions) $\vec J$, is the number of spin precessions 
around the $\hat n$ per turn on a trajectory,
viewed in a so--called {\em uniform precession frame} (UPF).
See \cite{beh2004} for 
precise definitions for smooth systems, i.e., systems with continuously differentiable
functions,  and for an explanation of how a particular ADST is, in fact, 
a member of an equivalence class. 
Note that although the systems in this paper are not smooth in $s$ due to the presence
of point--like snakes (see below), their smoothness in $\vec z$ facilitates a close analogy
with the smooth systems of \cite{beh2004}.

In general, an ADST does not exist if the trajectory is
on orbital resonance but on the other hand, one avoids running a
machine on orbital resonances, at least those of low order. If an ADST exists, it depends only on 
$\vec J$, hence the name ADST.

The ADST provides a way to quantify the degree of coherence between
the spin and orbital motion and thereby predict how strongly the electric 
and magnetic fields along particle trajectories disturb spins.
In particular, the spin motion can become very erratic close to the
{\em spin--orbit resonance} condition
$\nu_{\rm s} (\vec J) ~=~k_0 + k_{1} Q_{1} + k_{2} Q_{2} + k_{3} Q_{3}$
where the $Q$'s are orbital tunes and the $k$'s are integers. Near these resonances the ISF can spread out 
so that $P_{_{\rm lim}}$ is very small.
The spin tune on the design orbit $\nu_0 \equiv \nu_{{\rm s}}( \vec 0)$
always exists and so does $\hat n_0 (s) \equiv \hat n(\vec 0; s)$.

In this paper we shall be concerned mainly with those orbital resonances
where the $Q$'s are rational. We write the fractional parts, $[Q_i]$, of
rational tunes $Q_i$ ($i = 1,2,3$) as $a_i/b_i$ where the $a_i$ and
$b_i$ are integers.  Here and later the brackets $[...]$ are used to
signal the fractional part of a number.  For rational  $[Q_i]$ a trajectory is periodic
over $c$ turns where $c$ is the lowest common multiple of the $b_i$.
This opens the possibility that in this case the ISF at each $(\vec z;
s)$ can be obtained (up to a sign) as the unit length real eigenvector
of the $3\times 3$ orthogonal matrix representing the $c$--turn spin
map (c.f. the calculation of $\hat n_0$ from the 1--turn spin map on
the closed orbit).  However, the corresponding eigentune $c \nu_{\rm
  c}$ extracted from the complex eigenvalues $\Lambda_{\rm c} = e^{\pm
  2\pi i c \nu_{\rm c}}$, depends in general on the  synchrobetatron phases at
the starting $\vec z$. Thus in general $\nu_{\rm c}$ cannot be used to
find a spin tune.  Nevertheless if $c$ is very large the dependence of
$\nu_{\rm c}$ on the phases can be very weak so that it can
approximate well the ADST of nearby irrational tunes.  For
non--resonant orbital tunes, the spin tune can be obtained using the
SODOM--II algorithm \cite{ky99} or from averaging the {\em pseudo spin
  tune} \cite{gh2006,mv2000}.

In perfectly aligned flat rings with no solenoids, $\hat n_0$ is
vertical and $\nu_0$ can be chosen to be $a \gamma_0$ where $\gamma_0$
is the Lorentz factor on the closed orbit and $a$ is the gyromagnetic
anomaly of the particle.  In the absence of skew quadrupoles, the
primary disturbance to spin is then from the radial magnetic fields
along vertical betatron trajectories.  The disturbance can be very
strong and the beam polarisation can be small near the condition $a
\gamma_0 = \kappa \equiv k_0 \pm Q_2$ where $k_0$ is an integer and
mode 2 is vertical motion.  This can be understood in terms of the
``single resonance model'' (SRM) whereby a rotating wave approximation
is made in which the contribution to ${\vec \Omega}$ from the radial
field along a vertical betatron trajectory is dominated by the Fourier component at
$\kappa$ with {\em resonance strength} $\epsilon (J_2)$.  The SRM can
be solved exactly and the ISF is given by \cite{mane2} ${\hat
  n}(\phi_2) =  \pm  \left( \delta {\hat e}_2 + \epsilon
  ({\hat e}_1 \cos \phi_2 + {\hat e}_3 \sin \phi_2 ) \right)/ \lambda$
where $\delta = a \gamma_0 - \kappa$ is the distance in tune space to
the parent resonance, $\lambda = \sqrt {\delta^2 + \epsilon^2}$,
$\phi_2$ is the difference between the vertical betatron phase and the phase of the
Fourier component and $({\hat e}_1,{\hat e}_2,{\hat e}_3)$
are horizontal, vertical and longitudinal unit vectors.
The tilt of
$\hat n$ away from the vertical $\hat n_0$ is $| \arcsin (\epsilon
/\lambda) |$ so that it is $90^{\circ}$ at $\delta = 0$ for non--zero
$\epsilon$.  At large $|\delta |$, the equilibrium polarisation
directions $\hat n(J_2,\phi_2; s)$, are almost parallel to $\hat
n_0(s)$ but as we see from the above formula, at $\delta= 0$, $\hat n$
lies in the horizontal plane and $P_{_{\rm lim}} = 0$.  In this simple
model $\nu_{\rm s}$ exists and is well defined near spin--orbit
resonances for all $Q_2$. In our calculations we choose the phase of the Fourier harmonic
to be zero so that $\phi_2$ represents the phase of the vertical betatron motion.

It is found both in practice and in simulation, that in the
absence of special measures, acceleration of the beam through $\delta
= 0$ at practical rates can lead to loss of beam polarisation. This loss can be ascribed
to a loss of invariance of $J_{\rm s}$ and it can be
quantified in terms of the Froissart--Stora formula \cite{fs60}.
Luckily, the loss of polarisation can be reduced by installing pairs
of Siberian Snakes \cite{dk76,dk78},  magnet systems which rotate spins by $\pi$,
independently of $\vec z$, around a ``snake axis'' in the machine
plane.  For example, one puts two snakes at diametrically opposite
points on the ring.  Then ${\hat n_0 \cdot {\hat e}_2} = +1$ in one
half ring and $-1$ in the other.  With the snake axes relatively at
$90^{\circ}$, the fractional part of $\nu_0$ becomes $1/2$ for all
$\gamma_0$.  For calculations one often represents the snakes as
elements of zero length (``point--like snakes''). Then if, in addition,
the effect of vertical betatron motion is described by the SRM, and
orbital resonances are avoided, at most $J_2$, the fractional part of
the ADST is $1/2$ too, independently of $\gamma_0$
\cite{ky88,mane2001,spin2002}. This is a special feature of this model. Thus for $[Q_2]$  away from
$1/2$, the system is not at the first order spin--orbit resonance 
$\nu_{\rm s}(J_2) = [Q_2]$.
Therefore such resonances are not
crossed during acceleration through $\delta = 0$ and the polarisation
can be preserved. This is confirmed by tracking simulations.  
However, simulations have shown also that the
polarisation can still be lost if $[Q_2] = {\tilde a}_2/2 {\tilde b}_2$ 
where here, and later,
${\tilde a}_2$ and ${\tilde b}_2$ are odd positive integers with
${\tilde a}_2 < 2 {\tilde b}_2$ \cite{lt86,syl88,sylbook}.  This is
the ``snake resonance phenomenon'' and it has also had practical
consequences \cite{lt86,syl88,sylbook,ptitsyn2004,mbai2004}, especially for small ${\tilde
  b}_2$.  Such a $[Q_2]$ fits the condition $1/2 = (1 - {\tilde
  a}_2)/2 + {\tilde b}_2 [Q_2]$.  Since such tunes correspond to
orbital resonance an ADST does not exist at most amplitudes. Then,
according to our definition the system is not on a spin--orbit
resonance $\nu_{\rm s}(J_2) = (1 - {\tilde a}_2)/2 + {\tilde b}_2
[Q_2]$.  However, for nearby irrational $[Q_2]$ an ADST can exist, namely with the value 1/2.
Then one can say that the system is close to spin--orbit resonance.
This case is studied in Part II.
Because the system is on orbital resonance and using the analogy with the smooth systems \cite{beh2004}, 
even a smooth $\hat n$ need not be globally unique. 
Even if it were, there would be no guarantee that the maximum time
averaged polarisation on a torus would be given by $|\langle\hat
n(\vec z;s)\rangle|$.  We investigate these matters in the next
section.  Note that the rings in the Relativistic Heavy Ion Collider, RHIC \cite{ptitsyn2004,mbai2004}
contain two snakes and that the RHIC team has avoided running near snake--resonance 
vertical tunes.  Even away from the dangerous orbital
tunes just mentioned, snake layouts should be chosen carefully.
Methods for choosing layouts are discussed in \cite{gh2006,mv2000}.

Although one can describe spin motion in terms of orthogonal 3x3
matrices, here, we prefer to use SU(2) matrices.  Correspondingly, the
orientation of a spin is encoded in a two--component spinor\footnote{
   Of course, these spinors should not be interpreted as ``spin wave
    functions'': here we are dealing with classical
    equations of motion for spin expectation values.}.  
We write the SU(2) matrices as
\begin{eqnarray}
I \cos(\psi/2)  -i \vec {\sigma} \cdot \hat m \sin(\psi/2) 
\label{eq:2.1}
\end{eqnarray}
where $I$ is the $2\times2$ unit matrix, $\hat m$ is the
unit vector along the effective rotation axis, $\psi$ is the angle of
rotation around that axis and the three components of $\vec \sigma$
are the Pauli matrices. The rotation is right handed when $\psi > 0$.   
Equation (\ref{eq:2.1}) can be re--written as
\begin{eqnarray}
I r_0  -i \vec {\sigma} \cdot \vec r
\end{eqnarray}
where $\sum_{i=0}^3 r_i^2 = 1$. We call the real ordered quadruple $(r_0, \vec r)$
a {\em unit quaternion} \cite{ham1844,mv2000}. Spin maps are then concatenated using the
multiplication rule 
\begin{equation}
 (a_0,\,\vec a)\;(b_0,\,\vec b) =
 (a_0b_0-\vec a\cdot\vec b,\, a_0\vec b +\vec a b_0 +\vec a\times\vec b) =
 (c_0,\,\vec c)
\end{equation}
where $(a_0, \vec a)$, $(b_0, \vec b)$ and $(c_0, \vec c)$ are unit quaternions.
The elements of the usual $3\times3$ matrices are given
by $R_{ij} = (2r_0^2-1)\delta_{ij}+2r_ir_j+2r_0\epsilon_{ijk}r_k $ where $\delta_{ij}$ is 
the Kronecker symbol and $\epsilon_{ijk}$ is the Levi--Civita symbol. 
Note that the $R_{ij}$ are homogeneous quadratic forms in the $r_i$. This implies
that  $R_{ij}(r_0,\vec{r})=R_{ij}(-r_0,-\vec{r})$ which
simply reflects the fact that SU(2) covers SO(3) twice.
In this paper, as in \cite{spin2002}, we consider a system with two
point--like snakes placed at diametrically opposite points on the ring. The snake
axes are respectively at $0^{\circ}$ and $90^{\circ}$ to the
longitudinal direction.  The effect of vertical betatron motion is
modelled by the SRM.  The components of the unit quaternion for one
turn starting with  phase $\phi^0_2$ just before the first
($0^{\circ}$) snake are then
\begin{eqnarray}
r_0 &=&  \left(\frac{\epsilon}{\lambda}\right)^2 \,\sin^2\frac{\pi\lambda}{2}\,
         \sin(2 \phi^0_2 + 2\pi\kappa) \nonumber \\
r_1 &=&  \left( - \,\frac{\epsilon}{\lambda}\, \sin\pi\lambda \,\sin\pi\kappa 
                - 2\frac{\epsilon}{\lambda}\,\frac{\delta}{\lambda} \sin^2\frac{\pi\lambda}{2}                  \,\cos\pi\kappa
         \,\right)     \sin(\phi^0_2+\pi\kappa) \nonumber\\
r_2 &=& -\cos^2 \frac{\pi\lambda}{2} - \left(\frac{\delta}{\lambda}\right)^2 \, \sin^2\frac{\pi\lambda}{2}
        -\left(\frac{\epsilon}{\lambda}\right)^2 \, \sin^2\frac{\pi\lambda}{2} \, \cos(2 \phi^0_2+2\pi\kappa) \nonumber \\
r_3 &=& \left( - \, \frac{\epsilon}{\lambda} \,  \sin \pi\lambda \, \cos \pi\kappa 
               + 2 \, \frac{\epsilon}{\lambda} \,\frac{\delta}{\lambda}    \sin^2 \frac{\pi\lambda}{2} \,\sin\pi\kappa 
         \,\right)    \sin(\phi^0_2 +\pi\kappa) \, .
\label{eq:1.5}
\end{eqnarray}
As mentioned above, on orbital resonance, the vector $\hat n$ can be
obtained (up to a sign) as the eigenvector of unit length of the
appropriate $c$--turn spin map. In terms of unit quaternions,
$\hat n$ is simply the unit vector along the vector ${\vec r}^{(c)}$ 
for the $c$--turn unit quaternion and we are free to choose the sign.

It is clear  from (\ref{eq:1.5}) that with small but non--zero
$\epsilon/\lambda$, the 1--turn spin map is close to a
rotation by the angle $\pi$ around an axis close to the  vertical.  This is expected on
physical grounds too:  at large $|\delta|$, i.e., far from the parent
resonance, or at small $\epsilon$, the perturbation embodied in $\epsilon$ is relatively
unimportant and the spins precess by an amount per turn similar to
that on the design orbit. Then, the map for an odd number of turns is also
close to a rotation by the angle $\pi$ around the vertical but the
map for an even number of turns is close to the identity.
If $\lambda$ is an even integer, the 1--turn spin map is always a rotation 
by the angle $\pi$ around the vertical.  

It is straightforward to show that at most small values of
$\epsilon/\lambda$ and with $[Q_2] = a_2/b_2$, the rotation vector ${\vec r}^{(b_2)}$ for
a $b_2$--turn map is close to vertical for
 $[Q_2] = \;1/3$, $ \;2/3$, $ \;1/5$, $ \;2/5$, $ \;3/5$, $ \;4/5$, $ \;1/7$, $ \;2/7$, $ \;3/7$, $ \;4/7$, $ \;5/7$, $ \;6/7$, $ \dots$
and for $[Q_2] = \;1/4$, $ \;3/4$, $ \;1/8$, $ \;3/8$, $ \;5/8$, $ \;7/8$, $ \;1/12$, $ \;5/12$, $ \;7/12$, $ \;11/12$, $ \dots$,
 and that unless $\lambda$ is an even integer, it is close to the horizontal plane for
$[Q_2] = 1/6$, $ \;5/6$, $ \;1/10$, $ \;3/10$, $ \;7/10$, $ \;9/10$, $ \;1/14$, $ \;3/14$, $ \;5/14$, $
\;9/14$, $ \;11/14$, $ \;13/14$, $\cdots$, corresponding to snake resonances.

\section{Polarisation in the model ring at rational  $[Q_2]$}

We now use our model to study and contrast the equilibrium beam
polarisation, the time averaged beam polarisation and the beam
polarisation surviving after acceleration, for 
the first members of the three classes of rational tunes just listed,
namely for $[Q_2] = 1/3$, $ \;1/4$ and $1/6$.  We are primarily interested
in $[Q_2]$ at and near $1/6$ but the other cases serve to familiarise
the reader with the ``normal'' cases.

\subsection{Off orbital resonance}
To set the scene, and at variance with the title of this section, 
we first consider a case where the system is off
orbital resonance and off spin--orbit resonance so that the smooth ISF $\hat n$ is globally unique.
Thus figure 1 shows the components of $\hat n$ for $\delta = 0$ 
in the range $0 < [\phi_2/{2 \pi}] \le 1$ obtained by
stroboscopic averaging \cite{hh96,hvb99,gh2006,mv2000} at the irrational tune
$Q_2 = 47 + \sqrt{5} - 2 = 47.236067977\dots$\footnote{Of course, we are aware that in calculations in a digital computer, 
all irrational numbers must be represented by rational numbers, but then of very high order.}.
In this and in all other figures in this  paper, the spins are viewed just before the $0^{\circ}$ snake.
Furthermore, for all calculations in this paper, the resonance strength, $\epsilon$, is  0.4 and the integer $k_0$ is 1800, corresponding to a proton
energy of about 970 GeV.  These are the values used in \cite{spin2002}
and we use them again here to allow comparisons to be made.

\begin{figure}[htbp]
\begin{center}
\includegraphics[width=9.0cm,angle =0.]{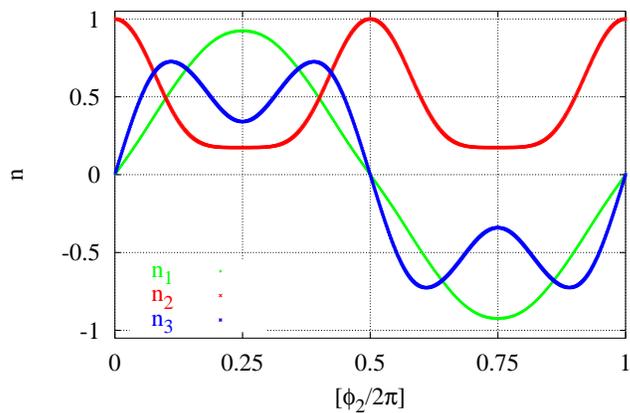}  
\caption{\footnotesize{The three components of ${\hat n}(\phi_2)$ 
for the SRM with  2 Siberian Snakes 
with axes at $0^{\circ}$ and $90^{\circ}$ and for $[Q_2] = 0.236067977\dots$. Viewing point: just before
the $0^{\circ}$ 
snake. $\delta = 0$ and  $\epsilon = 0.4$.
}}
\end{center}
\end{figure}
We remind the reader that
$\hat n$ is $2\pi$--periodic in $\phi_2$.  In principle, the
stroboscopic averaging could have been carried out at each value of
$[\phi_2/{2 \pi}]$ separately. However, away from orbital resonances
one can cover a torus by simply finding $\hat n$ at some $[\phi_2/{2 \pi}]$, 
setting a spin parallel to this $\hat n$
and then recording the spin components 
while transporting the spin for a large number of turns. 
Since $J_{\rm s}$ is invariant along a trajectory we then have the components of $\hat n$
all along the trajectory.  
This is the approach adopted for figure 1 and we see confirmation that 
$\hat n$ is a single valued continuous function of $[\phi_2/{2 \pi}]$.
The average $\langle\hat n\rangle$ 
of $\hat n$ over $\phi_2$ is vertical and $P_{_{\rm lim}} = 0.47$. The ADST  is $1/2$.  
\begin{figure}[htbp]
\begin{center}
\includegraphics[width=9.0cm,angle =0.]{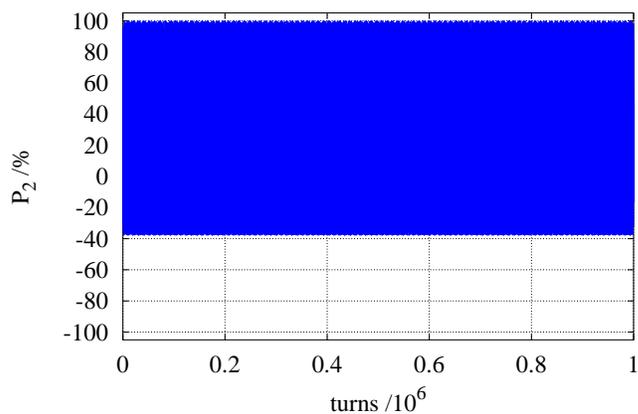}  
\caption{\footnotesize{For  initially
vertical spins, the vertical component of the beam polarisation, 
sampled every 100 turns,  at $\delta = 0$ for $[Q_2] = 0.236067977\dots$.
}}
\end{center}
\end{figure}

Figure 2 shows the beam polarisation, sampled every hundred turns for
$10^6$ turns, for an ensemble of particles distributed uniformly in
the range $0 < [\phi_2/{2 \pi}] \le 1$ at $\delta = 0$ when the spins
are all initially vertically upward. The horizontal components remain
at zero but the vertical component oscillates, at least for millions
of turns, between time independent maxima and minima with a time
average of about 0.3. As expected, this is less than $P_{_{\rm lim}}$.
A constant polarisation equal to the maximum 0.47 could have been
attained by setting the spins initially parallel to their respective
$\hat n$ vectors. See also figure 9 in \cite{hh96}. Inspection of the
turn--by--turn data reveals that the oscillations have a period of about four turns, as
expected for a $[Q_2]$ close to one quarter and an ADST of $1/2$.
In the simple SRM and at $\delta = 0$ the analogous simulation would 
exhibit a beam polarisation oscillating between $+1$ and $-1$
as the spins precessed around the horizontal $\hat n$ at a rate
$\lambda = \epsilon$.
\begin{figure}[htbp]
\begin{center}
\includegraphics[width=9.0cm,angle =0.]{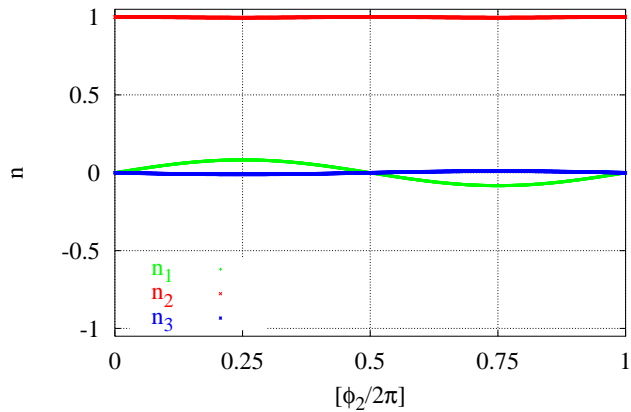}  
\caption{\footnotesize{The three components of ${\hat n}(\phi_2)$ 
at $\delta = 10.6$ for $[Q_2] = 0.236067977\dots$.
}}
\end{center}
\end{figure}

Figure 3 shows the components of $\hat n$ for the parameters of
figure 1 except with $\delta =10.6$, a value corresponding to  a beam energy far from
that of the parent resonance, with non--even $\lambda$ but otherwise arbitrary.  The
vectors ${\hat n} (\phi_2)$ are almost vertical so that $P_{_{\rm
    lim}}$ is high, namely 0.998.
\begin{figure}[htbp]
\begin{center}
\includegraphics[width=9.0cm,angle =0.]{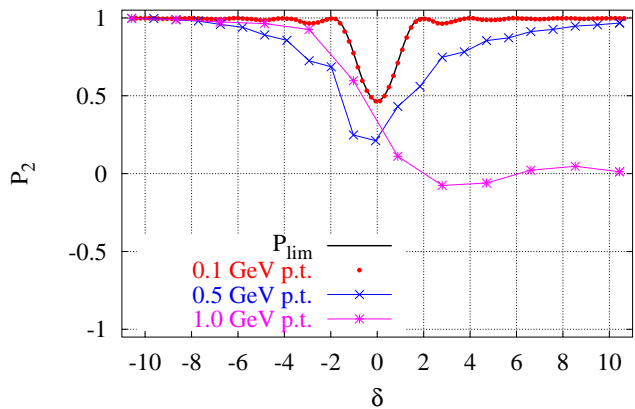}  
\caption{\footnotesize{With each spin initially parallel to its $\hat n$, the 
beam polarisation sampled turn--by--turn, for $[Q_2] = 0.236067977\dots$ 
during acceleration from $\delta = -10.6$ to $\delta = +10.6$ 
at the rates of 100 MeV, 500 MeV and 1 GeV per turn.
}}
\end{center}
\end{figure}
Figure 4 shows the curve for $P_{_{\rm lim}}$ together with the beam
polarisations, as ensembles are accelerated through $\delta = 0$ at
the rates of 100 MeV, 500 MeV and 1 GeV per turn (p.t.).  The
acceleration is simulated by incrementing $\delta$ by four equal
amounts, namely just after each snake and at the mid--points of the
two arcs.  At the start, $\delta = -10.6$ and the particles are
distributed uniformly in $[\phi_2/2 \pi]$  with each spin
initially set parallel to its corresponding ${\hat n}(\phi_2)$, which
is almost vertical.  For protons, a rate of 100 MeV per turn
corresponds to $\Delta \approx 0.19$ for the change of $a \gamma_0$ per
turn. For this rate the beam polarisation follows the curve for
$P_{_{\rm lim}}$ vs. $\delta$, dipping to the value 0.47 at $\delta =
0$.  Moreover, detailed inspection shows that at each $\delta$ the
distribution of spins matches the ISF.  This is  a nice demonstration
of the adiabatic invariance of $J_{\rm s}$ in this case
\cite{hde2006}.  The invariance of $J_{\rm s}$ is lost at the higher
rates. Slightly different curves are obtained if the spins are set 
vertically upward at the start. 

The rate of 100 MeV per turn corresponds to a value
${\epsilon}^2/\alpha \approx 5.3$ in the Froissart--Stora formula
\cite{fs60} where $\alpha = \Delta /{2\pi}$.  The Froissart--Stora
formula describes the final polarisation when a spin--orbit resonance
is crossed in the SRM and for these parameters it would predict almost
full spin flip, corresponding to adiabaticity.
However, our model includes the snakes and there are therefore no
first order spin--orbit resonances to cross. So the Froissart--Stora
formula does not apply.  Nevertheless for our model, the rate of 100 MeV
per turn is adiabatic.

\subsection{On orbital resonance: $[Q_2] =1/3$} 

We now consider our first case of orbital resonance, namely with $Q_2
= 47 + 1/3$, corresponding to odd $a_2$ and $b_2$.  Figure 5 shows
the components of $\hat n$ at $\delta = 0$ and $\epsilon = 0.4$.
These components are obtained by normalising to unity the ${\vec
  r}^{(3)}$ corresponding to three turns in the range $0 < [\phi_2/{2
  \pi}] \le 1/3$, namely $0^{\circ}$ to $120^{\circ}$, and then transporting  the $\hat
n$ for each $[\phi_2/{2 \pi}]$ in this range for two or more turns with the
1--turn spin map, thereby filling  up the full phase
range. Note that the curves are single valued functions of
$[\phi_2/{2\pi}]$ as required. The average $|\langle\hat n\rangle|$ 
of $\hat n$ over $[\phi_2/{2 \pi}]$ in
figure 5 is 0.05 and $\langle\hat n\rangle$ is vertical.
\begin{figure}[htbp]
\begin{center}
\includegraphics[width=9.0cm,angle =0.]{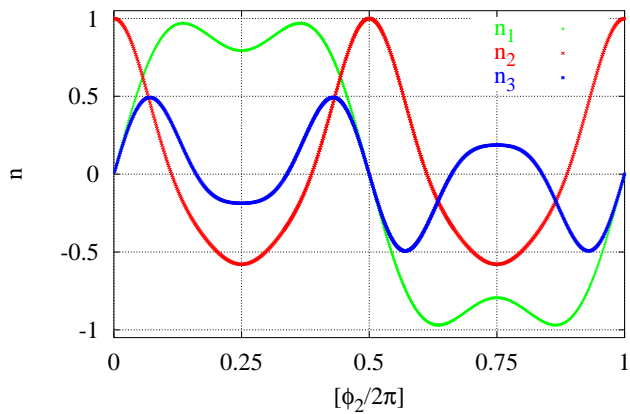}  
\caption{\footnotesize{The three components of ${\hat n}(\phi_2)$ 
at $\delta = 0$  for $[Q_2] = 1/3$.
}}
\end{center}
\end{figure}
While the smooth ISF $\hat n$ of figure 5 is globally unique, one looses global 
uniqueness if one allows discontinuities, as demonstrated in 
figure 6. There, we introduce  changes of sign in $\hat n$ by hand at the arbitrarily chosen
angles of $17.5^{\circ}$ and $90^{\circ}$, while constructing $\hat n$ in the range $0^{\circ}$ to $120^{\circ}$
using ${\vec r}^{(3)}$.
We then transport this $\hat n$ for two or
more turns as before. Naturally, the sign--discontinuities 
(often  simply called ``discontinuities'' from now on) 
are transported
too. In particular, we see that the transported $\hat n$ is still a single
valued function of $[\phi_2/2 \pi]$. 
The average $|\langle\hat
n\rangle|$ in figure 6 is 0.164. It is clear that neither $\hat n$ nor
$|\langle\hat n\rangle|$ are unique.  Of course, an unlimited number
of discontinuities could be introduced in the same way.  Then the
curves would be smooth almost nowhere.
Each of the $\hat n$ obtained in this way would
correspond to a permissible equilibrium spin distribution.
\begin{figure}[htbp]
\begin{center}
\includegraphics[width=9.0cm,angle =0.]{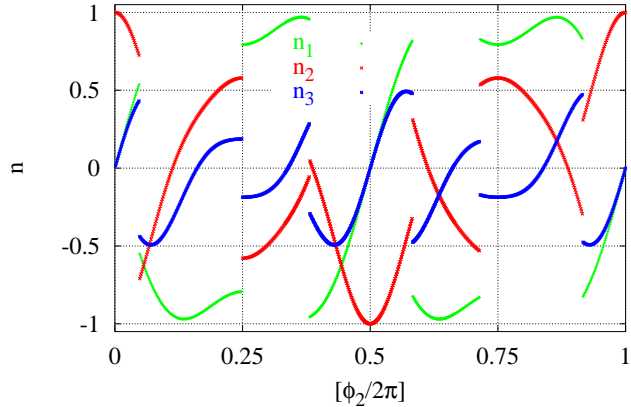}  
\caption{\footnotesize{The three components of ${\hat n}(\phi_2)$ 
at $\delta = 0$ for $[Q_2] = 1/3$. Sign--discontinuities have been introduced by hand.
}}
\end{center}
\end{figure}
The $\hat n$ obtained by stroboscopic averaging \cite{hh96} over the whole range
$0 < [\phi_2/{2 \pi}] \le 1$ can have discontinuities with positions
that depend on the ``seed'' spin field used in the stroboscopic
average but these discontinuities can be removed to give the curves in figure 5.
Since these discontinuities are sign discontinuities, we do not
exclude the possibility that the ISF is locally unique.  However, this
issue is beyond the scope of this paper since it would lead us into a
discussion of regularity conditions.
\begin{figure}[htbp]
\begin{center}
\includegraphics[width=9.0cm,angle =0.]{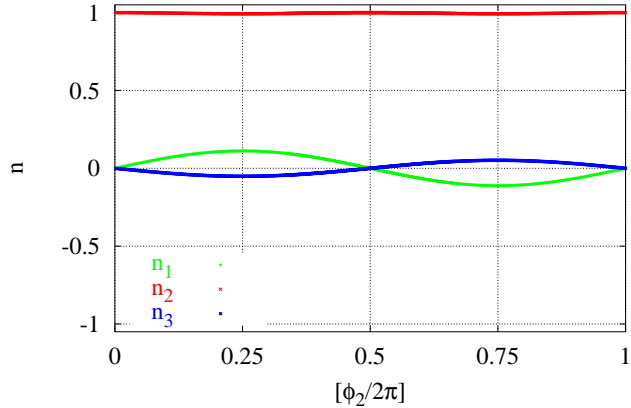}  
\caption{\footnotesize{The three components of ${\hat n}(\phi_2)$ 
at $\delta = 10.6$   for $[Q_2] = 1/3$.
}}
\end{center}
\end{figure}

If the long term tracking simulation of figure 2 is repeated but with
$[Q_2] = 1/3$, the vertical component of the beam polarisation oscillates quickly 
between about -0.3 and +0.8 for at
least $5\cdot 10^{6}$ turns with a time average of about 0.25. This is higher
than the $|\langle\hat n\rangle|$ in figure 5 but no significance can be
attributed to this since $|\langle\hat n\rangle|$ is not unique.

Figure 7 shows the components of the smooth ISF $\hat n$ for the conditions of figure 5
but with $\delta = 10.6$.
$|\langle\hat n\rangle|$ is high as expected, namely 0.997
since ${\vec r}^{(3)}$ is close to vertical. The existence of the $\hat n$ of figure 7, means that an ensemble of
exactly vertical spins is close to a permissible equilibrium spin
distribution.

Figure 8 shows the beam polarisation for acceleration 
through $\delta = 0$ from $\delta = -10.6$ to $\delta = +10.6$ at the rates of 50 MeV, 300 MeV and 1 GeV per turn 
for this $Q_2$.  At the start, the particles are 
distributed uniformly in  $[\phi_2/{2 \pi}]$ and
the spins are set parallel to
the almost vertical $\hat n$ vectors of the smooth ISF.
Up to an acceleration rate of 50 MeV per turn, $J_{\rm
  s}$ is invariant, with the beam polarisation dipping down to 0.05
 around $\delta = 0$ and returning to a high value at
the end.  This is a demonstration that with the chosen smooth $\hat n$,
$J_{\rm s}$ can be adiabatically invariant, although the proof in
\cite{hde2006} does not guarantee this because the system is on orbital
resonance.  At the higher acceleration rates, the invariance is lost.
By using stroboscopic averaging for irrational $[Q_2]$ near $1/3$ one finds ISFs 
similar to that in figure 5.
\begin{figure}[htbp]
\begin{center}
\includegraphics[width=9.0cm,angle =0.]{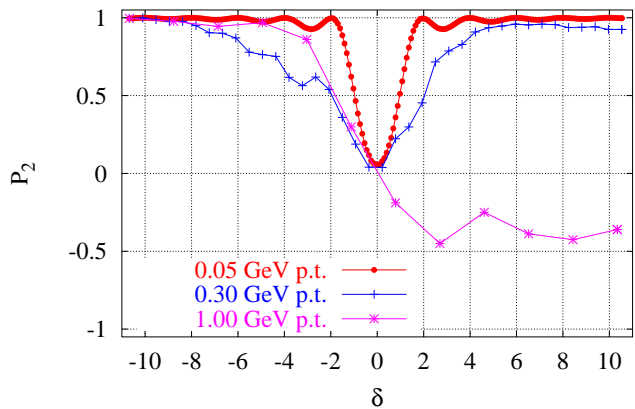}  
\caption{\footnotesize{With each spin initially parallel to its $\hat n$, 
the beam polarisation, sampled turn--by--turn, for $[Q_2] = 1/3$ 
during acceleration from $\delta = -10.6$ to $\delta = +10.6$ 
at the rates of 50 MeV, 300 MeV and 1 GeV per turn.
}}
\end{center}
\end{figure}

\subsection{On orbital resonance: $[Q_2] =1/4$} 

For our second case of orbital resonance we choose $Q_2 = 47 + 1/4$,
corresponding to an odd $a_2$ and a $b_2$ which is twice an even integer.
Figure 9 shows the components of $\hat n$ at $\delta = 0$ and $\epsilon
= 0.4$ obtained, in analogy with the previous case, from ${\vec
  r}^{(4)}$ in the range $0 < [\phi_2/{2 \pi}] \le 1/4$ and from
transporting those $\hat n$ for three or more turns.  In this case we
see ``stray'' points at multiples of $45^{\circ}$  corresponding to the phases
where the 4--turn map is the identity. For this figure we have imposed
the constraint that the components are continuous in the range $0^{\circ}$ to $90^{\circ}$, 
apart from the stray points. If we had not imposed smoothness,
the components would have changed sign at $45^{\circ}$ and the resulting
discontinuities would have been transported to the remainder of the
phase range.  So, for these parameters and for $[Q_2] =1/4$, $\hat n$
can have discontinuities as in the case of any rational $Q_2$. But in
contrast to a case discussed below, these discontinuities can be suppressed. 
The $\hat n$ obtained
by stroboscopic averaging over the whole range $0 < [\phi_2/{2 \pi}]
\le 1$ is smooth as in figure 9. Of course, as in the case of $[Q_2]
=1/3$, we can also introduce an unlimited number of sign--discontinuities.
The curves of figure 9 give $|\langle\hat n\rangle| = 0.43$.
\begin{figure}[htbp]
\begin{center}
\includegraphics[width=9.0cm,angle =0.]{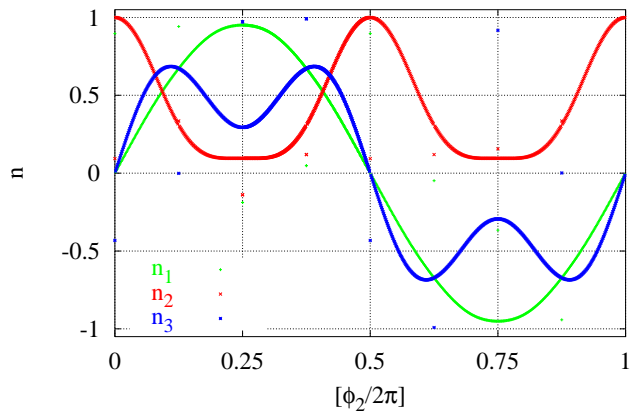}  
\caption{\footnotesize{The three components of ${\hat n}(\phi_2)$ 
at $\delta = 0$  for $[Q_2] = 1/4$.
}}
\end{center}
\end{figure}
Note the similarity between figure 9 and figure 1. Such similarities
are seen with other irrational $[Q_2]$ near 1/4 and indicate a weak
dependence of $\hat n$ on such irrational $[Q_2]$. This is consistent with the prediction
in \cite[Section 4.8]{mv2000} that in mid--plane symmetric rings the ISF is well behaved close to the condition
$\nu_0 = k_0 + 2 k_2 Q_2$, ($k_0, k_2 \in {\mathbb Z})$.

If the long term tracking simulation of figure 2 is repeated but with
$[Q_2] = 1/4$, the vertical component of the beam polarisation oscillates quickly, initially  between 
about -0.1 and +0.7.  But these limits gradually change and become  0.1 and 0.4 respectively after
$5\cdot 10^{6}$ turns. The time average of about 0.25. This is lower
than the $|\langle\hat n\rangle|$ in figure 9 but no significance can be
attributed to this since $|\langle\hat n\rangle|$ is not unique.

Figure 10 shows the components of $\hat n$ for the conditions of figure 9
but with $\delta = 10.6$. The average $|\langle\hat
n\rangle|$ is  0.99. Note that in contrast to the 3--turn map used for
$[Q_2] =1/3$, at large $|\delta|$ the 4--turn map is close to the
identity. Nevertheless, ${\vec r}^{(4)}$ is close to vertical.
\begin{figure}[htbp]
\begin{center}
\includegraphics[width=9.0cm,angle =0.]{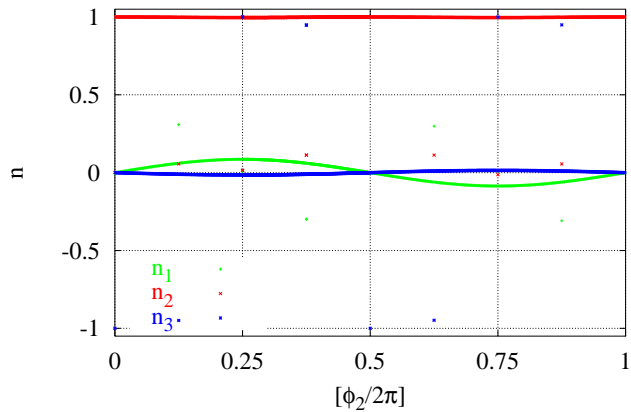}  
\caption{\footnotesize{The three components of ${\hat n}(\phi_2)$ 
at $\delta = 10.6$   for $[Q_2] = 1/4$.
}}
\end{center}
\end{figure}

Figure 11 shows the beam polarisation as the simulation of figure 4 is 
repeated for $[Q_2] =1/4$.  At the start, the spins are set parallel to
the almost vertical $\hat n$ vectors of the smooth ISF.
Up to an acceleration rate of 100 MeV per turn, $J_{\rm s}$ is
invariant, with the beam polarisation dipping down to 0.43
around $\delta = 0$ and returning to a high value at the
end.  This is again a demonstration that with the chosen $\hat n$,
$J_{\rm s}$ can be adiabatically invariant although the system is
on orbital resonance.  At the higher acceleration rates, the invariance is
lost.

\begin{figure}[htbp]
\begin{center}
\includegraphics[width=9.0cm,angle =0.]{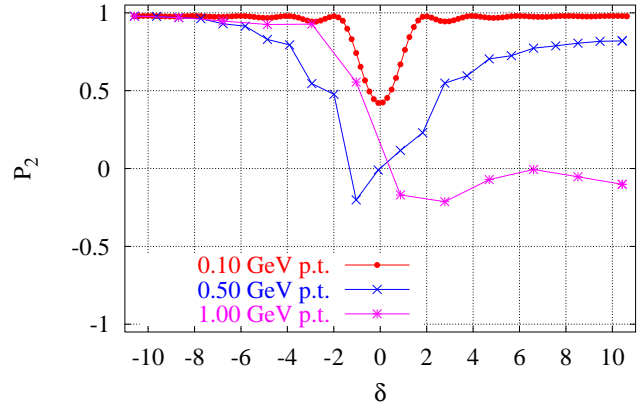}  
\caption{\footnotesize{With each spin initially parallel to its $\hat n$, 
the beam polarisation, sampled turn--by--turn, 
for $[Q_2] = 1/4$ during acceleration from
$\delta = -10.6$ to $\delta = +10.6$ at the rates of 100 MeV, 500 MeV and 1 GeV per turn.
}}
\end{center}
\end{figure}

\subsection{On orbital resonance: $[Q_2] =1/6$} 

We now come to the first of the two cases of primary interest for this study, namely the 
case when $[Q_2] =1/6$, i.e., a case of a snake 
resonance. Again, the integer part of $Q_2$ is 47 and  $\epsilon= 0.4$.  
Figure 12 shows the components of
$\hat n$ at $\delta = 0$ obtained 
by transporting for five or more turns the $\hat n$ obtained
from ${\vec r}^{(6)}$ in the range
$0 < [\phi_2/{2 \pi}] \le 1/6$.
\begin{figure}[htbp]
\begin{center}
\includegraphics[width=9.0cm,angle =0.]{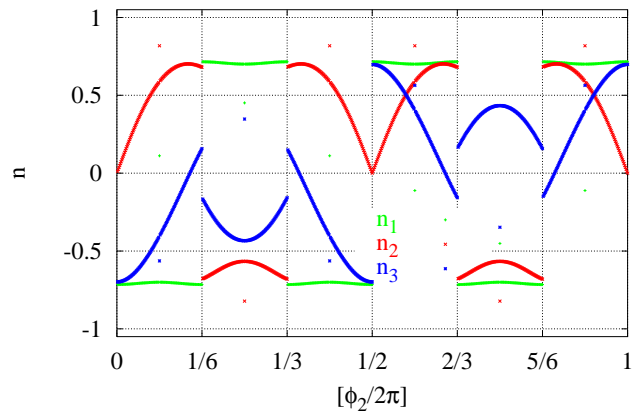}  
\caption{\footnotesize{The three components of ${\hat n}(\phi_2)$ 
at $\delta = 0$  for $[Q_2] = 1/6$.
}}
\end{center}
\end{figure}

We see stray points at phases which are multiples of $30^{\circ}$ and $90^{\circ}$
corresponding to the phases where the 6--turn map is the identity.
The vector ${\vec r}^{(6)}$ has sign--discontinuities at these points
but for this figure we have imposed the constraint that the components
of $\hat n$ are continuous in the range $0^{\circ}$ to $60^{\circ}$,
apart from the stray points.  One sees that $\hat n$ still has
discontinuities, namely at phases which are multiples of $60^{\circ}$. Thus, in spite
of smoothing $\hat n$ in the initial range of $0^{\circ}$ to
$60^{\circ}$, discontinuities persist. They cannot be removed without
creating a vector field which becomes double valued when it is
transported turn--by--turn.  However, the discontinuities can be moved. 
These effects  explain
the failure of the MILES algorithm for $\hat n$ at snake--resonance
tunes in \cite{mane2004} where the need for discontinuities in this model is
nevertheless disputed. It is clear that the curves in figs.\,7 and 8
in \cite{mane2004} do not represent $\hat n$ \cite{spin2002}.

Stroboscopic averaging over the whole range $0 < [\phi_2/{2 \pi}] \le
1$ generates the curves of figure 12 directly i.e., without extra
smoothing. The discontinuities of $\hat n$ occur at phases where the
raw stroboscopic average passes through zero.  The passage through
zero is smooth. So discontinuities in $\hat n$ do not imply
discontinuities in the stroboscopic average.

Our numerical calculations show that $\hat n$ has such
discontinuities at snake--resonance tunes at most values of
$\epsilon$ and that the minimum number of discontinuities is $2 {\tilde b}_2$.

Of course, if $\hat n$ is represented as the locus
of points on the  unit 2--sphere, one finds disjoint
segments.  The average $|\langle\hat n\rangle|$ over $[\phi_2/{2\pi}]$ 
in figure 12 is 0.13. An arbitrary number of extra discontinuities 
can be introduced by hand.

If the long term tracking simulation of figure 2 is repeated but with
$[Q_2] = 1/6$, the polarisation oscillates quickly, but with constant upper and
lower limits with a time average of about 0.1, at least up to $5\cdot
10^{6}$ turns. 
Thus the time averaged polarisation does not vanish. 
This is illustrated in figure 13. 
\begin{figure}[htbp]
\begin{center}
\includegraphics[width=9.0cm,angle =-0.]{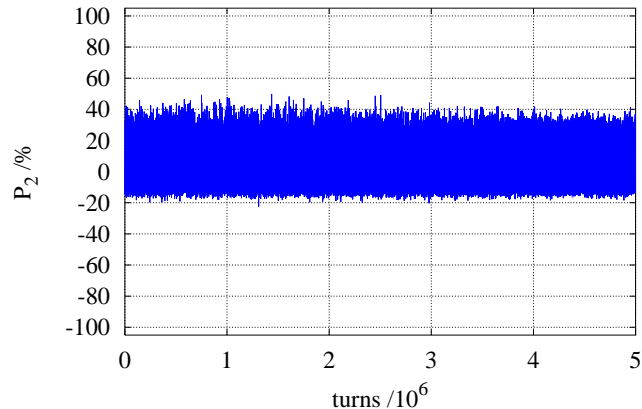}  
\caption{\footnotesize{For  initially
vertical spins, the vertical component of the beam polarisation, sampled every 1000 turns,
at $\delta = 0$ for $[Q_2] = 1/6$.
}}
\end{center}
\end{figure}

\begin{figure}[htbp]
\begin{center}
\includegraphics[width=9.0cm,angle =0.]{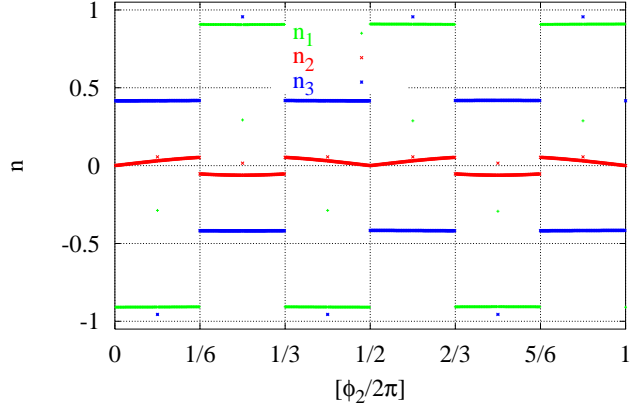}  
\caption{\footnotesize{The three components of ${\hat n}(\phi_2)$ 
at $\delta = 10.6$ for $[Q_2] = 1/6$.
}}
\end{center}
\end{figure}

Figure 14 shows $\hat n$ obtained as for figure 12 but with $\delta =
10.6$.  Except when $\lambda$ is an even integer this is typical of
the $\hat n$ at large $|\delta|$ (and also at small $\epsilon$).  
The value of ${\vec r}^{(6)}$ is
very small and the 6--turn spin map is close to a rotation of $2\pi$
around $\hat n$.  The discontinuities persist but in contrast to the
earlier examples, the vertical component of $\hat n$ is close to zero
and the horizontal components are piece--wise almost independent of
$[\phi_2/{2 \pi}]$.  The average $|\langle\hat n(\phi_2)\rangle|$ is essentially zero. It
would remain close to zero if sign--discontinuities were introduced by
hand.  Since the horizontal components of $\hat n$ are piece--wise
almost independent of $[\phi_2/{2 \pi}]$ but also different, and since the
1--turn spin map is a rotation of about $\pi$ around an axis close to the vertical, it
essentially changes their signs from turn to turn, causing the
discontinuities.  Such discontinuities do not occur at large $|\delta|$
for $[Q_2] =1/3$ or $[Q_2] =1/4$ in figures 7 and 10 because $\hat n$ is close to
vertical.
The curves of figure  14 deform continuously into those of figure 12
as $\delta$ is reduced to zero. The analogous curves for the other three tunes
show the same kind of behaviour and, of course, that behaviour is a prerequisite  
for $J_{\rm s}$ is to be invariant in figures  4, 8 and 11.

For $[Q_2] = 1/6$ with $\epsilon = 0.4$ and large non--even integer
$\lambda$, all equilibrium spin distributions have spins close to the
horizontal plane. Thus a spin distribution in which all spins are
initially vertically upward cannot be in equilibrium. This is confirmed in
figure 15 where we repeat the long term tracking simulation of figs.\,2
and 13 but at $\delta = -10.6$ and $[Q_2] =1/6$.  We now see that the
polarisation falls, but slowly, over many tens of thousands of
turns and subsequently oscillates around zero. Then the time averaged
polarisation is close to $|\langle\hat n(\phi_2)\rangle| \approx 0$.
Nevertheless, since the system is on orbital resonance, the theorem
\cite{gh2006,mv2000} on the maximum time averaged polarisation does
not enforce this.
\begin{figure}[htbp]
\begin{center}
\includegraphics[width=9.0cm,angle =0.]{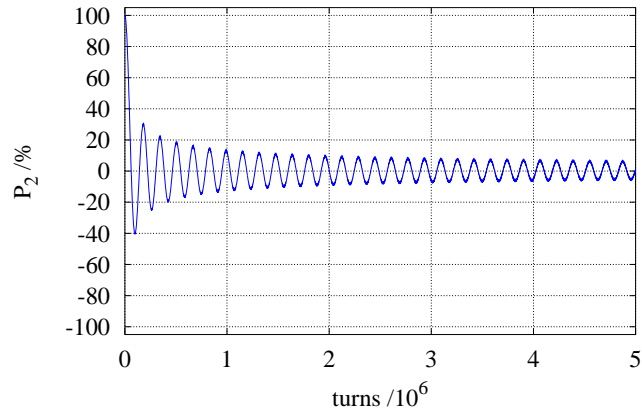}  
\caption{\footnotesize{For  initially
vertical spins, the vertical component of the beam polarisation, sampled every 1000 turns,
at $\delta = -10.6$ for $[Q_2] = 1/6$.
}}
\end{center}
\end{figure}
\begin{figure}[htbp]
\begin{center}
\includegraphics[width=9.0cm,angle =-0.]{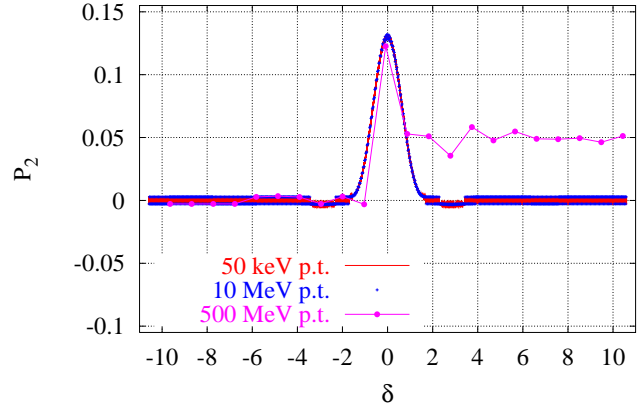}  
\caption{\footnotesize{The beam polarisation for $[Q_2] = 1/6$ during acceleration from
$\delta = -10.6$ to $\delta = +10.6$ at the rates of 50 KeV, 10 MeV and 500 MeV per turn
with the spins initially parallel to $\hat n$.  
}}
\end{center}
\end{figure}
\begin{figure}[htbp]
\begin{center}
\includegraphics[width=9.0cm,angle =0.]{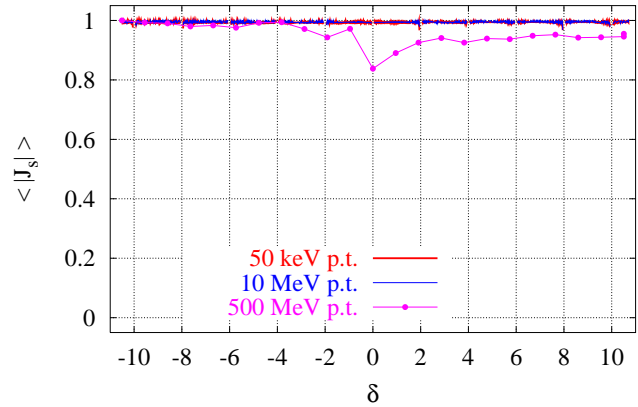}  
\caption{\footnotesize{With each spin initially parallel to its $\hat n$, 
$\langle J_{\rm s} \rangle$ during acceleration from $\delta = -10.6$ to $\delta = +10.6$ 
at the rates of 50 KeV, 10 MeV and 500 MeV per turn with $[Q_2] = 1/6$.
}}
\end{center}
\end{figure}
\begin{figure}[htbp]
\begin{center}
\includegraphics[width=9.0cm,angle =0.]{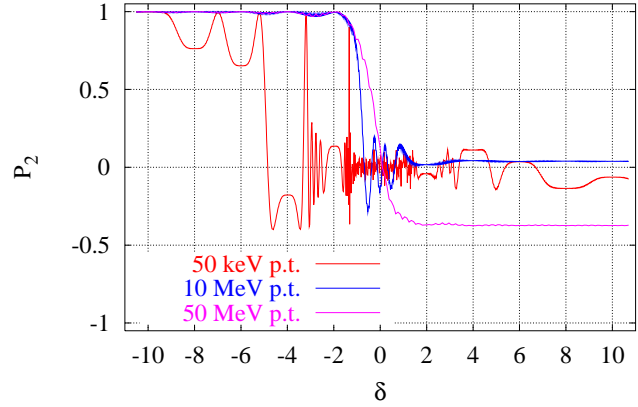}  
\caption{\footnotesize{For initially vertical spins, the beam polarisation
for $[Q_2] = 1/6$ during acceleration from
$\delta = -10.6$ to $\delta = +10.6$ at the rates of 50 KeV, 10 MeV  and 50 MeV per turn.
}}
\end{center}
\end{figure}
\begin{figure}[htbp]
\begin{center}
\includegraphics[width=9.0cm,angle =0.]{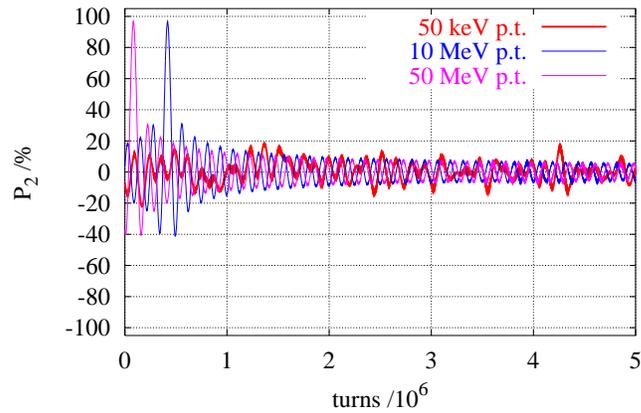}  
\caption{\footnotesize{The beam polarisation  
for $[Q_2] = 1/6$ when  $\delta$ is frozen at  $+10.6$ after the acceleration 
cycle of figure 18, and the spins are tracked for a further  $5 \cdot 10^6$ turns.  
}}
\end{center}
\end{figure}
Although the initial spin distribution is not in equilibrium, it is
not surprising that it takes about $10^5$ turns before the
polarisation reaches zero.  This is due to the fact that at large $|\delta|$ the
eigentune, $6 \nu_{6}$, of the 6--turn spin map is almost
independent of $[\phi_2/{2 \pi}]$ and very close to an integer for this
case.  Since $J_{\rm s}$ is invariant along a trajectory, we can view
the motion of a spin as a precession at a fixed angle ${\cos}^{-1}
(\vec S \cdot \hat n /|\vec S|)$ around its $\hat n$.  In this case
the angles are about $ 90^{\circ}$. With eigentunes almost independent
of $[\phi_2/{2 \pi}]$ and close to an integer, the projections of spins on the
planes perpendicular to their respective $\hat n$'s spread out
(decoher) only slowly.  Then, at the viewing position, the spins return
almost to their original directions after six turns.

For large $|\delta|$, the 1--turn spin map corresponds to a
rotation of about $\pi$ around an axis close to the vertical.  So, it is again no
surprise that the polarisation in figure 15 takes many turns to reach
zero.  For even larger $|\delta|$ (e.g., over 100), $\hat n$ can be
taken to be horizontal but the polarisation remains vertical and it
takes many millions of turns for it to show signs of falling.  There
is no fall if $\lambda$ is an even integer since then, the 6-turn map
is the identity.

Figure 16 shows the beam polarisation for acceleration 
through $\delta = 0$ at the rates of 50 KeV, 10 MeV and 500 MeV per turn 
for $[Q_2] =1/6$.  At the start, the particles are uniformly
distributed in $[\phi/{2 \pi}]$ and the spins are set
parallel to the almost horizontal $\hat n$ vectors of  that  ISF which deforms into the ISF's of figures 12 and 14.
The initial beam polarisation is essentially zero.  During
acceleration at rates up to 50 KeV per turn, the beam polarisation
{\em rises} to $0.13$, corresponding to the $|\langle\hat n\rangle|$
of figure 12, and then returns to around zero.  
A detailed inspection of the data shows that for 
a rate of  10 MeV per turn, the spins deviate slightly from their respective $\hat n$ vectors
at large $|\delta|$. However, this effect is not apparent in the average over  
$[\phi_2/{2 \pi}]$ contained in the beam polarisation.
This is again a
demonstration that with the chosen $\hat n$ and the chosen layout of accelerating cavities, $J_{\rm s}$ can be
approximately invariant even for these discontinuous ISF's and that at the higher
acceleration rates, the invariance is lost.  The approximate  invariance
is confirmed in figure 17 which shows the corresponding behaviour of the phase average of  $J_{\rm s}$,
$\langle J_{\rm s} \rangle$. 
In figure 17 we have suppressed data at $\delta$'s where
$\hat n$ is indeterminate because $\lambda$ is an even integer.

Figure 18 shows the beam polarisation as the simulation of figure 16
is repeated but with the spins initially vertically upward and 
for rates of 50 KeV and 10 MeV per turn and for 50 MeV per turn, where $J_{\rm s}$ is still approximately
invariant.  For these rates of 
acceleration the angle between a spin and its $\hat n$ remains around  $90^{\circ}$.
Then the beam polarisation during acceleration depends
just on the geometry of the ISF and on the history of the rate of
decoherence of the projections of the spins on the planes perpendicular to the $\hat
n$'s. These rates depend, in turn, on the magnitude of $6 \nu_{6}$ and
its dependence on $[\phi_2/{2 \pi}]$.  We therefore expect that the final
polarisation could depend sensitively on the magnitude of the rate of acceleration
and on its time dependence.
This is confirmed in
figure 18 which shows that at a rate of 50 KeV per turn, the
polarisation is effectively lost at positive $\delta$ but that at the
much higher rate of 50 MeV per turn the final polarisation is around
-0.4 at the end of the acceleration cycle.    By now, the reader will have realised that 
the polarisation of -0.4 cannot represent an equilibrium state.
This is confirmed in figure 19 where, after acceleration up to $\delta = 10.6$,
$\delta$ is frozen and the ensembles are tracked for a further $5 \cdot 10^6$
turns.  Figure 19 shows that after some large oscillations the polarisation gradually  
decays to zero in a way and on a time scale familiar from figure 15.
It also shows that although the polarisation can be small at the end of the acceleration
(as in the case of 10 MeV/turn), the spin distribution is by no means isotropic but is such that 
the polarisation can return to a large value later. 
In fact after the $5 \cdot 10^6$ turns, the curves of spin vector versus $[\phi_2/2\pi]$ are 
smooth curves for all three acceleration rates\footnote{
 This vindicates the advice in \cite[Section I]{beh2004} on the use of the term ``depolarisation''.}.
This suggests that contrary to conventional
expectation, a complete loss of polarisation is not inevitable during
acceleration exactly at a snake resonance with $[Q_2] = 1/6$, at least not within the
confines of our model.  
This completes Part I of our investigation.

\section{Summary and conclusion}
In this paper we have presented and contrasted four scenarios for spin
motion on and off orbital resonance within the confines of our simple
model, and by this means we have developed a clean, elegant account of
the special features of spin motion at a snake resonance.  In all
four cases $\vec S \cdot \hat n$ is an invariant at low enough rates
of acceleration.  For the first three cases ($[Q_2] =
0.236067977\dots, \;1/3, \;1/4$) the ISF is close to vertical at large
$|\delta|$, i.e., far away from the energy for the parent resonance,
and the spin motion is unexceptional. For example, after acceleration from a large negative 
$\delta$ to a high positive $\delta$, an initially vertical spin is still close to vertical. 
These cases serve to emphasise
the exceptional form of the ISF when $[Q_2] = 1/6$. In this case, far
away from the parent resonance, the ISF lies close to the horizontal
plane. Then in contrast to the other three cases, an ensemble of
particles with a uniform distribution of $[\phi_2/2 \pi]$ and with vertically upward
spins, cannot be at spin equilibrium. The subsequent evolution
of the beam polarisation depends on the chosen initial $\delta$ and is
exemplified in figures 13 and  15.  In particular, the polarisation
oscillates at a rate depending on the proximity of the eigentune of
the 6--turn spin map to an integer and on the extent of the variation
of that eigentune with $[\phi_2/{2 \pi}]$. Then at the energy of the parent
resonance ($\delta = 0$), the polarisation oscillates quickly and the
time averaged polarisation is small but non--zero.  At most large
$|\delta|$, the time averaged polarisation is zero but the
polarisation oscillates slowly and it reaches zero for the first time
only after many thousands of turns.

As soon as one sees that at most large $|\delta|$ the ISF for $[Q_2] =
1/6$ lies close to the horizontal plane, it is no surprise that 
in this case the
time averaged beam polarisation can become small in the long term.
Acceleration adds little to the story, except that
within our model,  after starting with an ensemble of vertical spins at $\delta = -10.6$, 
the final polarisation depends on the rate at which one passes from the 
spin motion underlying figure 15 to the spin motion underlying figure 13 and then beyond
to large positive $\delta$.
The key features of spin motion at $[Q_2] = 1/6$ are encoded in the ISF.
We see no necessity to invoke the perturbed spin tune \cite{syl88,sylbook}.
Instead, we appeal to the eigentune of the 6--turn spin map, a quantity with  
physical significance.

We emphasise that the main results presented here refer to a very
special case, namely for our model right at $[Q_2] = 1/6$ and with
$\epsilon = 0.4$.  As pointed out in \cite{spin2002}, the ISF is
extremely complicated for values of $[Q_2]$ just below and just above
$1/6$.  
This is consistent with the prediction
in \cite[Section 4.8]{mv2000} that in mid--plane symmetric rings the ISF need not be  
well behaved close to the condition
$\nu_0 = k_0 + (2 k_2 + 1) Q_2$,  ($k_0, k_2 \in {\mathbb Z}$).
Thus in Part II of this study we extend our calculations to
cover such values of $[Q_2]$ and to larger values of $\epsilon$.  It
will be shown there that although the ISF for $[Q_2] = 1/6$ has the
special form described above, this is an exception and that the loss
of polarisation during acceleration near to $[Q_2] = 1/6$ has a
different origin. We also comment on the findings in  \cite{ky88,buon85,ptitsin952}.

The analysis should then be extended to real synchrobetatron motion
with misalignments for a typical optic of a real ring and with the fields of real snakes.  See, for
example, \cite{ran2003}.  Other snake--resonance tunes should also
be covered.  We note with interest that according to simulations for
RHIC, the loss of polarisation during acceleration is less severe when
the simulations are carried out with the magnetic fields of real
snakes rather than with  point--like snakes
\cite{xk2003}.

\section*{Acknowledgements}
We thank K. Heinemann, G. H. Hoffstaetter and  J.A. Ellison for useful discussions and for valued
collaboration and we thank L. Malysheva for help during the preparation of this paper.

\end{document}